\documentclass[journal,11pt,draftclsnofoot,onecolumn]{IEEEtran}
\usepackage{cite}
\usepackage{graphicx}
\usepackage{psfrag}
\usepackage{url}
\usepackage{stfloats}
\usepackage{amsfonts,amssymb,amsmath,bm,paralist,theorem,cite,ifthen,color,amsmath}
\usepackage{array}
\usepackage{caption}
\usepackage{calc}
\usepackage{longtable}
\usepackage{subfigure}
\usepackage{mathrsfs}
\usepackage{epsfig}

\newtheorem{lemma}{Lemma}

\newtheorem{thm}{Theorem}

\theorembodyfont{\rmfamily}

\newcommand{\rank}{{\rm rank}}
\newcommand{\tr}{{\rm Tr}}
\newcommand{\st}{{\rm s.t.}}
\newcommand{\mbQ}{{\mathbf{Q}}}
\newcommand{\mbW}{{\mathbf{W}}}
\newcommand{\bW}{{\bf W}}
\newcommand{\bC}{{\bf C}}

\newcommand{\bv}{{\textbf{v}}}
\newcommand{\bH}{{{H}}}
\newcommand{\bG}{{G}}

\begin{document}
%
\title{Linear Transceiver Design for Interference Alignment: Complexity and Computation\thanks{This work is supported in part by the Army Research Office, Grant No. W911NF-09-1-0279, by the National Science Foundation, grant number
CMMI-0726336, and by a research gift from Huawei Technologies Inc.}}
%

\author{Meisam Razaviyayn, Maziar Sanjabi and Zhi-Quan Luo~\IEEEmembership{Fellow,~IEEE}
\thanks{The authors are with the Department of Electrical and Computer Engineering, University of Minnesota, 200 Union
Street SE, Minneapolis, MN 55455. Emails: \{razav002,sanja006,luozq\}@ece.umn.edu.}}

\markboth{IEEE Transactions on Information Theory~(submitted)}%
{IEEE Transactions on Information Theory~(submitted)} 
%
\maketitle
%
\begin{abstract}
Consider a MIMO interference channel whereby each transmitter and
receiver are equipped with multiple antennas. The basic problem is
to design optimal linear transceivers (or beamformers) that can
maximize system throughput. The recent work \cite{Jafar1} suggests
that optimal beamformers should maximize the total degrees of
freedom and achieve interference alignment in high SNR. In this
paper we first consider the interference alignment problem in
spatial domain and prove that the problem of maximizing the total
degrees of freedom for a given MIMO interference channel is
NP-hard. Furthermore, we show that even checking the achievability
of a given tuple of degrees of freedom for all receivers is
NP-hard when each receiver is equipped with at least three
antennas. Interestingly, the same problem becomes polynomial time solvable
when each transmit/receive node is equipped
with no more than two antennas. Finally, we propose a distributed algorithm for transmit
covariance matrix design, while assuming each receiver uses a
linear MMSE beamformer. The simulation results show that the
proposed algorithm outperforms the existing interference alignment
algorithms in terms of system throughput.
\end{abstract}
\newpage
\section{Introduction}
\label{sec:intro}

Consider a multiuser communication system in which a number of users
must share common resources such as frequency, time, or space. The
mathematical model for this communication scenario is the well-known
\textit{interference channel}, which consists of
multiple transmitters simultaneously sending messages
to their intended receivers while causing interference
to each other. Interference channel is a generic model for
multiuser communication and can be used in many practical
applications such as Digital Subscriber Lines (DSL) \cite{DSL},
Cognitive Radio (CR) systems \cite{CR} and ad-hoc wireless
networks \cite{AdHoc1,AdHoc2}.

A central issue in the study of interference channel is how to mitigate multiuser interference.
In practice, there are several commonly used methods for dealing with interference.
First, we can treat the interference as noise and just focus on extracting the desired signals (see \cite{IC7}, \cite{IC14}).
This approach is widely used in practice because of its simplicity and ease of implementation, but is known to be non-capacity achieving in general. 
An alternative technique is channel orthogonalization whereby transmitted signals are chosen to be nonoverlapping either in time, frequency or space, leading to Time Division Multiple Access, Frequency Division Multiple Access or Space Division Multiple Access respectively. While channel orthogonalization effectively eliminates multiuser interference, it can lead to inefficient use of communication resources and is also generally non-capacity achieving. Another interference management technique is to decode and remove interference. Specifically, when interference is strong relative to desired signals, a user can decode the interference first, then subtract it from the received signal, and finally decode its own message (see \cite{IC3} and \cite{IC3half}). This method is less common in practice due to its complexity and security issues.

In a cellular system, multi-cell interference management is a major challenge. So far
various base station cooperation techniques have been proposed to mitigate inter-cell
interferences, including multi-point coordinated transmission,
or network MIMO transmission \cite{co1,co3,co2}. Most of these techniques require each base station
to have full/partial channel state information (CSI) as well as the knowledge of actual
independent data streams to all remote terminals. With the complete sharing of
data streams and CSI, the multi-cell scenario is
effectively reduced to a single cell interference management
problem with either total \cite{[5]} or per-group-of-antennas
power constraints \cite{[6],[7]}. While these techniques can offer significant improvement on data throughput, they also have several drawbacks including stringent requirement on base station coordination, the large demand on the communication
bandwidth of backhaul links, and the heavy
computational load associated with the increasing number of cells \cite{[8],[9]}.

Theoretically, what is the optimal interference management strategy? The answer is related to the characterization of capacity region of an  interference channel, i.e., determining the set of {\color{black} rate tuples} that can be achieved
by the users simultaneously. For the noiseless case, the capacity region and the optimal precoding strategy of the two user interference channel is discussed in \cite{IC3} and \cite{IC12}.  In spite of intensive research on this subject over the past three decades (\cite{IC12} - \cite{IC13}), the capacity region of interference channels is still unknown for general case (even for small
number of users).
The lack of progress to characterize the capacity region for a
MIMO interference channel has motivated researchers to derive
various approximations of the capacity region. For example, the
maximum total degrees of freedom (DoF) corresponds to the first
order approximation of sum-rate capacity of an interference
channel at high SNR regime. Maximizing this approximation of sum-rate leads us to the interference alignment method \cite{Jafar1}.
For frequency selective channels, interference alignment corresponds to correlated signalling across different {\color{black} frequency} tones. This linear transceiver scheme for interference alignment is a generalization of the standard OFDMA scheme whereby each data stream is transmitted on a single subcarrier, which corresponds to using {\color{black} the standard} unit basis vectors $\{e_i\}$ (the $i$-th {\color{black} standard unit} vector) for transmit beamforming. 
The linear transceiver structure for interference alignment is more general since it does not require diagonal structure nor mutual orthogonality (two transmit covariance matrices $X$, $Y$ are said to be orthogonal if $\tr(XY)=0$).

If we remove mutual orthogonality condition and impose only
diagonal structure on transmit covariance matrices, then the interference management problem is reduced
to the dynamic spectrum management problem \cite{complexity} where the
goal is to find the optimal power allocation (i.e., optimal
diagonal transmit covariance matrices) which can maximize system
throughput. This problem has recently been a topic of intensive
research in the signal processing and communications communities. For diagonal matrix
channel case (e.g. frequency selective scenario), the problem of
maximizing sum-rate has been shown to be NP-hard
\cite{complexity}. Several algorithms have been proposed which
provide varied  performance in different channel conditions.
These include: Iterative Waterfilling Algorithm IWFA \cite{IWFA1},
Successive Convex Approximation Low complExity (SCALE) algorithm
\cite{scale}, Autonomous Spectrum Balancing (ASB)\cite{ASB},
Optimal Spectrum Balancing (OSB) \cite{OSB}. Furthermore,
different algorithms are proposed for the case when the channel
matrices are non-diagonal. Authors in \cite{IWFA2,IWFA3} proposed
IWFA based algorithms for power allocation. However, these selfish
approaches work well only in low SNR cases  or when the
interference is low.

Compared to the networked MIMO approach, interference alignment requires less information exchange among
transmitters, and is therefore simpler to implement in practice.
Recently two iterative algorithms have been proposed for interference alignment \cite{Jafar2,Heath}.
Both appear to work well in simulation of small systems (e.g., three users, each
equipped with two antennas).\footnote{Even though the two
algorithms were motivated from different perspectives, they are in
fact algorithmically identical.}  These algorithms require system
users to first specify the DoFs for all receivers and then attempt
to achieve them by iteratively aligning the interferences.
However, these algorithms can not check if a given {\color{black} tuple of DoF is}
achievable, nor is there  any guarantee for reaching interference
alignment even when the given {\color{black} tuple of DoF is} achievable. Moreover, by
focusing only on high {\color{black} SNR regime} and interference alignment, these
algorithms do not attempt any power allocations across different
data streams. This can result in linear transceivers with
suboptimal performance at low to intermediate SNRs.

In this paper, we consider the problem of maximizing the sum
of DoFs and the problem of checking if a given set of {\color{black} DoFs is} achievable with linear transceivers. 
We study the complexity status of both of these problems over the spatial domain and establish their NP-hardness.
These results suggest that the two existing algorithms for interference alignment \cite{Jafar2,Heath}
cannot converge in general. We also propose
a distributed algorithm to design linear transceivers for interference channels. Our approach is based on
using MMSE receivers while optimizing transmit covariance matrices for all transmitters. We maximize the weighted sum of a utility of SINR's for each data stream and use iterative convex optimization/relaxation to compute a (local)
optimal solution. The utility function is {\color{black}
${\rm SINR}/(1+{\rm SINR})$
}
which converges to $1$ when {\color{black} SINR $\to\infty$}, and
is proportional to SINR when the {\color{black} SINR value is small}. In this way, maximizing the sum of utilities for all data streams corresponds
to maximizing the total DoF when the noise vanishes.
Simulations show that our algorithm performs well in all SNR regions and can
deliver far superior sum-rate performance than the existing interference alignment algorithms of
\cite{Jafar2,Heath}.
Compared to the networked MIMO approach which requires sharing of data streams,
our linear transceiver design algorithm requires
less information exchange: at each iteration, only small covariance matrices
are exchanged, the size of which are proportional to the antenna numbers at
each transmitter or receiver.


%

\section{System Model}
\label{sec:System_Model} Consider a $K$-user MIMO interference
channel with $K$ transmitter - receiver pairs. Let
$\mathbf{H}_{kj}$ be an $N_k \times M_j$ matrix representing
the channel gain from transmitter~$j$ to receiver~$k$, where $M_j$
and $N_k$ denote the number of antennas at transmitter~$j$ and
receiver~$k$, respectively. The received signal at receiver~$k$ is
given by
\begin{align}
 \mathbf{y}_k = \sum_{j=1}^{K} \mathbf{H}_{kj} \mathbf{x}_j
+ \mathbf{n}_k,\nonumber
\end{align}
\normalsize where $\mathbf{x}_j$ is an $M_j \times 1$ random
vector that represents the transmitted signal of user~$j$ and
$\mathbf{n}_k \sim \mathcal{N} (\mathbf{0}, \sigma^2 \mathbf{I})$
is a zero mean
additive white Gaussian noise.\\
For practical considerations, we focus on optimal {\em linear}
transmit and receive strategies that can maximize system
throughput. In particular, suppose transmitter~$k$ uses a beamforming
matrix $\mathbf{V}_k$  to send the signal vector
$\mathbf{s}_k$ to its intended receiver~$k$. At the receiver side,
receiver~$k$ estimates the transmitted data vector $\mathbf{s}_k$
by using a linear beamforming matrix $\mathbf{U}_k$, i.e.,
\begin{align}
 \mathbf{x}_k = \mathbf{V}_k \; \mathbf{s}_k, \;\;\;\;\;\;\;
\hat{\mathbf{s}}_k = \mathbf{U}_k^T \mathbf{y}_k,\nonumber
\end{align}
\normalsize where the data vector $\mathbf{s}_k \in
\mathbb{C}^{d_k \times 1}$ is normalized so that $E[\mathbf{s}_k
\mathbf{s}_k^T] = \mathbf{I}$, and $\hat{\mathbf{s}}_k$ is the
estimate of $\mathbf{s}_k$ at $k$-th receiver. $\mathbf{V}_k \in
\mathbb{C}^{M_k \times d_k}$ and $\mathbf{U}_k \in \mathbb{C}^{N_k
\times d_k}$ are the beamforming matrices at the transmitter and
the receiver of user~$k$, respectively.

It is known that the problem of designing optimal beamformers to
maximize sum-rate of the system is NP-hard \cite{complexity} even
in the single transmit/receive antenna case. Notice that recent
works \cite{Jafar1,Jafar2} suggest that the optimal strategy
should have interference alignment structure in the high SNR
regime. Therefore, we are led to find a linear
transmission-reception strategy that can maximize the total
degrees of freedom. In the next section, we provide the complexity analysis of this problem.

\section{NP-Hardness of Optimal Interference Alignment}
\label{sec:NP_Hardness}

In this section, 
we show that for a given channel, not only the problem of finding
the maximum DoF is NP-hard, but also the problem of checking the
achievability of a given tuple of DoF, $(d_1,...,d_K)$, is NP-hard
when there are at least 3 antennas at each node.

Notice that the interference alignment conditions in the $k$-th
receiver are
\begin{align}
\small
&\mathbf{U}_k^T\mathbf{H}_{kj}\mathbf{V}_j=0,\;\;\;\;\  \forall j\neq k, \label{eq:First_IA_Condition}\\
&\rank\left(\mathbf{U}_k^T \mathbf{H}_{kk}\mathbf{V}_k\right)=d_k. \label{eq:Second_IA_Condition}
\end{align}
\normalsize The first equation guarantees that all the
interference is in the subspace orthogonal to $\mathbf{U}_k$ while
the second one assures that the signal subspace $\mathbf{H}_{kk}
\mathbf{V}_k$ has dimension $d_k$ and is linearly independent of
the interference subspace.

{\color{black}In the sequel,} we examine the solvability of above interference alignment problem \eqref{eq:First_IA_Condition} - \eqref{eq:Second_IA_Condition} in two different cases.
\begin{thm} \label{thm:NPhard}
For a $K$ user MIMO interference channel, maximizing the total
DoF, namely,
\begin{align}
\small
\max_{\{\mathbf{U}_k,\mathbf{V}_k\}_{k=1}^K} \;\;\ &\sum_{k=1}^K{d_k}\nonumber\\
\st \;\;\;\;\;\;\;\ &\mathbf{U}_k^T\mathbf{H}_{kj}\mathbf{V}_j =
0,\;\;\;\;\;\;k=1,..,K,
 \;\;j\neq k \nonumber\\
&\rank\left(\mathbf{U}_k^T\mathbf{H}_{kk}\mathbf{V}_k\right)=d_k,\;\;\;\;\;\;k=1,..,K
 \nonumber
\end{align}
\normalsize is NP-hard. Moreover, if each node is equipped with at
least 3 antennas, then the problem of checking the achievability
of a given tuple of DoF, $(d_1, d_2, \ldots, d_K)$, is also
NP-hard.
\end{thm}
\begin{proof}
The proof of the first part is based on a polynomial time
reduction from the maximum independent set problem which is known
to be NP-complete. For a given arbitrary graph $\bG=(V,E)$, where
$|V|=K$, consider a $K$ user interference channel that each
receiver and transmitter  has a single antenna. Moreover, the
channel coefficients are given by:
\begin{align}
h_{jk} = \left\{
\begin{array}{cc}
1, & \text{if}\;\;j = k \;\;\text{or}\;\; (k,j) \in E , \\
0, & \text{otherwise}.\\
\end{array}
\right.
 \nonumber
\end{align}
\normalsize It can be checked that the receiver nodes can only
achieve a DoF of either 0 or 1, and those receiver nodes achieving
a DoF of 1 form an independent set in $G$. Thus, the problem of
maximizing the total DoF for the above interference channel is
equivalent to the problem of finding the maximum independent set
of vertices in the graph $G$.\\
In order to prove the second part we use a polynomial reduction from the
3-colorability problem. The latter problem is to determine whether
the nodes of a graph can be assigned one of the three possible
colors so that no two adjacent nodes are colored the same. The
3-colorability problem is known to be NP-Complete. There are two
main steps in the construction. In the first step, some dummy
nodes are added to the channel in order to force a discrete
structure such that each non-dummy node may only have one of the
three possible cases. The second step is to define the direct
channels in order to make a polynomial reduction from the
3-colorability of an arbitrary graph to this problem.

For an arbitrary graph $G$ with $N$ nodes, we will construct a special MIMO interference channel for which the achievability of one degree of freedom at each user is equivalent to  the 3-colarability of $G$.
In our construction, the MIMO interference channel will have two types of users: $N$ main users, each equipped with 3 antennas at their transmitters and receivers and $11N$ dummy users which will be defined later. Hence the total number of users is $12N$. In the rest of the proof we suppose that each user ({\color{black} either the dummy user or the main user}) wants to send one data stream. In other words we want to check if the tuple of all ones is achievable by the constructed interference channel or not.

We divide the dummy users into two groups. The number of dummy users in the first group is $2N$ and the number of dummy users in the second one is $9N$. Each dummy user in the first group has 3 antennas at its receiver and transmitter, while each dummy user in the second group has two antennas at its transmitter and receiver.
Let us further arrange the $2N$ dummy users in the first group into $N$ subsets each containing two users. We denote these subsets as ${A}_i, i=1,...,N,|{A}_i|=2$. We also denote the users in the set ${A}_i$ as $a_{i,1}$ and $a_{i,2}$, and associate them to the $i$-th main user. For notational consistency, we denote main user $i$ as $a_{i,0}$. We will also use $a_{i,k,j}$ to denote the $j$-th transmit antenna of user $a_{i,k}$, where $1\le i\le N$, $k=0,1,2$ and $j=1,2,3$.
Similarly, we partition the set of $9N$ dummy users in the second group
into $N$ subsets ${B}_i, i=1,...,N$, each containing exactly 9 dummy users denoted by $b_{i,\ell}$, with $\ell=1,..,9$.
Each of these $9$ dummy users will have two receiving antennas which we denote as $b_{i,\ell,m}$, with $m=1,2$.

Now for any fixed $i$ and $j$, we consider any size-2 subset of $\{a_{i,k,j}:k=0,1,2\} $, e.g., $\{a_{i,0,j},a_{i,1,j}\}$. For each fixed {\color{black} $i$ and $ j$}, there are exactly 3 of these cardinality-2 subsets. Since there are 3 different choices of $j$, we have a total of 9 subsets of this kind for any fixed $i$. Let us index these 9 subsets by $\ell,\ \ell=1,..,9$, and assign the $\ell$-th subset to user $b_{i,\ell}$ in ${B}_i$.
Now we define the links in the channel for the users in ${A}_i$ and ${B}_i$. First, the channel matrices of all the direct links for any of the dummy users are $\textbf{I}$ (where $\textbf{I}$ is the identity matrix of the appropriate size). In addition, none of the dummy users in ${B}_i$ ($i=1,2,...,N$) cause interference to the other users (which means that the channel gains between their transmit antennas and the other users' receive antennas are {\color{black} all zero}). Now for the aforementioned $\ell$-th subset which we denote as ${S}_{i,\ell}=\{a_{i,k_{\ell_1},j_{\ell_1}},a_{i,k_{\ell_2},j_{\ell_2}}\}$, we connect  $a_{i,k_{\ell_1},j_{\ell_1}}$ and $a_{i,k_{\ell_2},j_{\ell_2}}$ to $b_{i,\ell,1}$ and to $b_{i,\ell,2}$, respectively. Here by connecting a transmit antenna to a receive antenna we mean that the channel coefficient between these two antennas is $1$. This situation is shown in the figure \ref{FIG:example} for the case ${S}_{i,1}=\{a_{i,0,1},a_{i,1,1}\}$. Furthermore, we assume that dummy users $a_{i,k},\;k=1,2$  do not suffer from any  interference.
\begin{figure}[htb] 
   \centering
   \includegraphics[width=7cm]{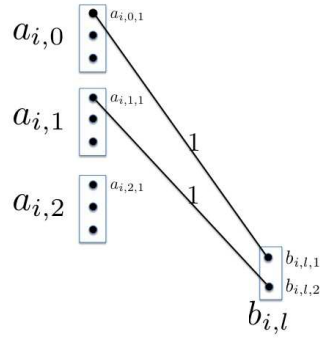}
   \caption{Channels to the dummy receiver ${\color{black} b_{i,\ell}}$}
   \label{FIG:example}
\end{figure}

Suppose that user $a_{i,k}$ ($k=0,1,2$) uses the transmit beamforming vector $(v_{i,k,1},v_{i,k,2},v_{i,k,3})$. Then the interference received at the dummy receiver of $b_{i,\ell}$ will be:
\begin{align}
\textbf{I}_{b_{i,\ell}}=(v_{i,k_{\ell_1},j_{\ell_1}}s_{i,k_{\ell_1}},v_{i,k_{\ell_1},j_{\ell_2}}s_{i,k_{\ell_2}})
\end{align}
where $s_{i,k}$ is the signal user $a_{i,k}$ intends to send.
Notice that the signals which two different users want to transmit are statistically independent. As a consequence, if we want to have interference alignment at the receiver of $b_{i,\ell}$, so that this user can send its own data stream, it is necessary and sufficient {\color{black} to have  $v_{i,k_{\ell_1},j_{\ell_1}} v_{i,k_{\ell_2},j_{\ell_2}}=0$.} Hence, having the interference alignment at $b_{i,\ell}$ for all $\ell=1,..,9$ is equivalent to the fact that users $a_{i,k},k=0,1,2$ cannot send their messages through the antennas with the same index, simultaneously. For example, if $v_{i,0,1}\neq0$ then $v_{i,1,1}$ and $v_{i,2,1}$ have to be zero. On the other hand, considering the fact that each user needs to send one data stream, it follows that none of the users $a_{i,k},\ k=0,1,2,$ can send their message on two of their antennas simultaneously, {\color{black} because otherwise if for example $a_{i,0}$ sends its message on two antennas, then it would result in insufficient spatial dimension for either $a_{i,1}$ or $a_{i,2}$.}


As an immediate consequence of these two facts we have just mentioned, we can conclude that the transmit beamforming vector at each user $a_{i,k},k=0,1,2,$ must be proportional to one of the vectors $[1,0,0]^T$, $[0,1,0]^T$ or $[0,0,1]^T$. This is true specially for the main user $i$. As we are not concerned about the constant factors, we have successfully imposed a discrete structure on the problem solution so far. Notice that each dummy user $b_{i,\ell}$ has a total of 2 dimensions in its receiver. Since we have aligned the interference at each dummy user $b_{i,\ell}$, these users can communicate their data streams easily along the remaining dimension left for them in their receivers and remove interference which lies in the other dimension. 
Moreover, since in our construction the dummy users $a_{i,k},\;k=1,2$  do not experience any interference from other users and their direct channel is $\textbf{I}$, so these users can easily achieve one degree of freedom. 
Thus, we only need to take care of  the main users.

For each of the $N$ main users, we must pick one of the three transmit beamforming vectors $[1,0,0]^T$, $[0,1,0]^T$ or $[0,0,1]^T$ in order to achieve interference alignment at all the main receivers. We suppose all the direct channels for the main users, $\textbf{H}_{ii}$, are $\textbf{I}$. For the cross channels, we use the structure of graph $G=({V},{E})$.
 For each edge $(i,j)$ in $G$, we set $\textbf{H}_{ij}=\textbf{H}_{ji}=\textbf{I}$. Otherwise we set $\textbf{H}_{ij}=\textbf{H}_{ji}=\textbf{0}$ (zero matrix of appropriate size). Consequently, the main users $i$ and $j$ interfere with each other if and only if they are connected to each other in graph $G$. 
We claim that achieving interference alignment in the above MIMO interference channel is equivalent to  3-colorability of graph $G$. This is because each user can choose 3 possible beamforming vectors, each corresponding to a different color. If main user $i$ chooses one of the three possible beamforming vectors (or one of the three colors), then this beamforming vector cannot be chosen by any other main users adjacent to the main user $i$ in the graph $G$, otherwise the interference would appear in the desired signal space at the receiver of main user $i$. This establishes the equivalence between the 3-colorability of $G$ and the achievability of one degree of freedom for each user in the constructed MIMO interference channel. Since 3-colorability problem is NP-hard, it follows that the problem of checking the feasibility of interference alignment is also NP-hard.
\end{proof}

Theorem \ref{thm:NPhard} shows that the problem of checking the achievability of a given tuple of DoF is NP-hard if all users (or at least a constant fraction of them) are equipped with at least three antennas. Our next result shows that when each user is equipped with no more than two antennas, the same problem can be solved in polynomial time. To this end, we need to define some notations and make some observations. First of all, the interference alignment problem is equivalent to finding the signal subspaces at the transmitters and the interference subspaces at the receivers such that the interference alignment conditions are satisfied, i.e.,
\begin{align}
&d_k = \dim (\mathcal{S}_k)\nonumber\\
&\mathbf{H}_{kk} \mathcal{S}_k\;   \underline{\perp} \; \mathcal{I}_k \nonumber\\
&\mathbf{H}_{kj} \mathcal{S}_{j}\;\subseteq \;\mathcal{I}_k\;\quad \forall j \neq k, \nonumber
\end{align}
where $\mathcal{S}_k$ and $\mathcal{I}_k$ denote the signal
subspace at the transmitter $k$ and the interference subspace at
receiver $k$, respectively. The operator $\underline{\perp}$ represents the
linear independence of two subspaces. The first condition
implies that the signal space has dimension $d_k$ while the second
condition says that the interference subspace and the received
signal subspace must be linearly independent. Finally, the third
condition assures that the interference from other users lies in
the interference subspace (which is linearly independent of the signal subspace).

Notice that in the 2-antenna case, if $d_j = d_k = 1$ and
$\rank(\mathbf{H}_{kj}) = 2$, and the interference subspace $\mathcal{I}_k$ is known, then
$\mathcal{S}_j$ can be uniquely determined by $\mathcal{S}_j =
\mathbf{H}_{kj}^{-1} \mathcal{I}_k$, for any $j\neq k$. Conversely, if
$\mathcal{S}_j$ is known, we can uniquely find the
interference subspace of user $k$, i.e., $\mathcal{I}_k =
\mathbf{H}_{kj} \mathcal{S}_j$. Thus, by starting from a node with a known subspace and traversing the interference links with full rank channel matrices, we can uniquely determine the signal subspaces in the transmitter sides and the interference subspaces at the receiver sides as long as they all have one DoF. Furthermore, if we find a loop of full rank interfering links, the signal subspaces at these nodes must be the eigenvector of the composite channel matrix of the corresponding loop. To make this point clear, consider a 4-user interference channel. If all interfering links are full rank, by starting from transmitter 1 and use the loop Tx1 $\rightarrow$ Rx2 $\rightarrow$ Tx3 $\rightarrow$ Rx4 $\rightarrow$ Tx1, we have the following relations
\begin{align}
\mathcal{I}_2 = \mathbf{H}_{21} \mathcal{S}_1 , \quad \mathcal{S}_3 = \mathbf{H}_{23}^{-1} \mathcal{I}_2, \quad \mathcal{I}_4 = \mathbf{H}_{43} \mathcal{S}_3, \quad \mathcal{S}_1 = \mathbf{H}_{41}^{-1} \mathcal{I}_4. \nonumber
\end{align}
 Thus, $\mathcal{S}_1$ must be the eigenvector of the loop channel matrix $\mathbf{H}_{41}^{-1} \mathbf{H}_{43} \mathbf{H}_{23}^{-1} \mathbf{H}_{21}$.
Using this observation and the idea of traversing the full rank interfering channel links, we can establish the polynomial solvability of the problem of checking the achievability of a given tuple of DoF.
\begin{thm}\label{thm:2}
For a $K$-user MIMO interference channel where each
transmit/receive node is equipped with at most two antennas, the
problem of checking the achievability of a given tuple of DoF is
polynomial time solvable.
\end{thm}
\begin{proof}
By assigning zero channel weight if necessary, we can assume without loss of generality that all transmitters/receivers are equipped with exactly two antennas, i.e., $M_k = N_k = 2 $, for all $k=1,2, \cdots, K$. Furthermore,
notice that if a user has zero DoF ($d_k=0$), then we can assign the zero beamforming vector to this user and remove it (both its transmitter and receiver) from the system. Thus, we can assume $1\le d_k\le 2$ for all  $k=1,2, \cdots, K$. {\color{black} We further assume that all the direct channel matrices $\mathbf{H}_{kk}, k =1,2,\ldots,K$, are nonzero}. Now the problem is to determine whether the given tuple of DoF $(d_1,d_2, \cdots, d_K)$ is achievable or not. To this end, we need to define two bipartite graphs over the nodes of the interference channel (one side of the graph consists of transmit nodes and the other consists of the receive nodes). In particular, we construct a bipartite graph $G$ by connecting the transmit node of user $i$ to the receive node of user $j$ if and only if the channel between them is nonzero, i.e., $\mathbf{H}_{ji} \neq 0$. Furthermore, we construct a bipartite subgraph $G' = (V',E') $ of $\bG$ by considering only the full rank links of $G$, i.e., connecting transmit node $i$ to the  receive node $j \neq i$ if and only if $\rank (\mathbf{H}_{ji}) =2$. Notice that the link between transmit node $i$ and receive node $i$ is not included in $G'$ even if $\rank (\mathbf{H}_{ii}) =2$.\\

{\color{black}In what follows, we first consider a simple case which gives us the idea of how a loop of rank 2 interfering channels forces a discrete structure on the choice of signaling subspaces at the transmitters. Then, using this idea, we provide the proof for the general case.\\}


{\color{black}Consider a connected component $H$ of $G$ where all the interfering links are full rank and connected, i.e., the induced subgraph of $H$ over $G'$ is connected and contains all the interfering links of $H$.
We first argue that
$H$ can not contain the receive node of any user $k$ with $d_k=2$. Suppose the contrary. Then 
the direct channel matrix, $\mathbf{H}_{kk}$, must be full rank. [If
$\mathbf{H}_{kk}$ is rank deficient, then the received signal subspace at receiver $k$ has dimension at most 1, which would make it impossible to achieve $d_k=2$.] We further claim that $\bH$ cannot contain any other nodes. Since the direct link between the transmit and receive nodes of user $k$ is not contained in $H$, it follows that the receive node of user $k$ must be connected to another transmit node $a$ in $\bH$. Let this node $a$ be associated with a user $j$ ($j\neq k$). Notice that user $j$ achieves a DoF at least 1 (since all zero DoF users have been removed from $\bG$). By definition, node $a$ must be connected to the receive node of user $k$ via a full rank cross talk channel matrix $\mathbf{H}_{kj}$. Thus, user $j$ will cause a nonzero interference subspace to user $k$, contradicting $d_k=2$. Since all users with DoF =0 has been removed from graph $G$,  we must have $d_k=1$ for all receive nodes in $H$. For the other case where node $a$ is a receive node of user $j$, then $a$ is linked to the transmit node of user $k$ via a full rank channel matrix. In this case, user $k$ will cause a 2-dimensional interference subspace to user $j$, making it impossible to have $d_j\ge 1$. }

We now assume that all receive nodes in $\bH$ have one DoF. 
We can start from an arbitrary initial node of $\bH$
and use Breadth First Search (BFS) to find a spanning tree.
Since each user has one DoF, the signal and interference spaces of all receive
nodes in $\bH$ are uniquely determined by the signal (or interference) space of the initial node. Since the initial node is arbitrary, this shows that the signal/interference spaces for all nodes in $\bH$ are linearly related to each other (via some constant composite channel matrices, see the discussion before Theorem 2). Fixing any one uniquely determines the rest. For the remaining edges (or links) not in the spanning tree, they each create a unique loop in the tree. We can compute the composite channel matrices for these loops (see the discussion before Theorem~\ref{thm:2}). Notice that each loop matrix (size $2 \times 2$) has either one, two or infinitely many eigenvectors (when the composite channel matrix is a constant multiple of identity matrix). Suppose a loop matrix (starting from a given transmit node, say $b$, in the loop) has one or two unique eigenvectors, then the signal space of node $b$ must be generated by one of these eigenvectors. In fact, since the beamforming vectors of nodes in $\bH$ are linearly related, each loop in $\bH$ places a restriction on the choice of beamforming vector of node $b$. Thus, for any fixed transmit node $b$ in $\bH$, there are multiple restriction sets, each corresponding to a loop in $\bH$ caused by adding an edge to the minimum spanning tree and each containing one/two one-dimensional subspaces from which node $b$'s signal space can be chosen. The receive nodes in $\bH$ can achieve interference alignment if and only if  these restricted sets of one-dimensional signal subspaces for node $b$ share a common one-dimensional subspace. Moreover, to ensure each user in $\bH$ achieves one DoF, we need to additionally make sure that the resulting interference subspaces at all receive nodes in $H$ are linearly independent from the corresponding respective signal subspaces. Since the total number of restriction sets is at most linear in the number of edges in $H$ and each restriction set contains at most two one-dimensional subspaces, checking if these restrictions have any common one-dimensional subspace can be carried out in $O(K^2)$ time. Moreover, for each common one-dimensional subspace, checking if the linear independence between the resulting signal subspace and interference subspace (already aligned) at each receive node can also be performed in time that is linear in the number of nodes in $H$, or in $O(K)$ time. 

{\color{black}Now we are ready to look into the general case in which the rank 1 links are considered as well as the full rank links. Since there is no interfering link between different connected components of $G$, we can assign the signal subspace for each connected component separately. Notice that the number of connected components of $G$ is at most $K$, we only need to assign transmit subspaces for every connected component of $G$ in polynomial time.\\

Let $\bH$ be a connected component of $\bG$. Let $\bH'\subseteq \bG'$ be a subgraph of $\bH$ which contains only links with full rank channel matrices. $\bH'$ can be decomposed into various connected components of $G'$. By the argument above for such components, the signal/interference spaces for the nodes in these connected components (consisting of at least two nodes) can be assigned in one of the two ways:}
\begin{enumerate}
\item [(B1)]
The connected component contains a cycle with a channel matrix that is not equal to a constant multiple of the identity matrix. In the case, the beamforming vectors of all nodes can be determined from the eigenvector(s) of a certain loop channel matrix. In this case, there are at most two possible choices of signal/interference space for each node.
\item [(B2)]The connected component has no loops (i.e., forms a tree) or if every loop has a composite channel matrix that is a constant multiple of the identity matrix. In this case, the signal/interference spaces of all nodes are linearly related to one another. The signal/interference space of one node can be fixed at an arbitrary one-dimensional subspace. Once this is fixed, the signal/interference spaces of other nodes can be derived uniquely.
\end{enumerate}

Consider a rank-1 interfering link in $\bH$ with channel matrix $\mathbf{H}_{ij}$ ($i\neq j$). If user $j$ transmits in the null of $\mathbf{H}_{ij}$, then the signalling subspace of user $j$ is known, i.e., $\mathcal{S}_j = \textit{Null} (\mathbf{H}_{ij})$. Otherwise, the interference subspace at user $i$ is known, i.e., $\mathcal{I}_i = \textit{Range} (\mathbf{H}_{ij})$. This is because $d_i\ge 1$, so we have $\dim {\cal I}_i\le 1$. This plus the fact that $\textit{Range} (\mathbf{H}_{ij})\subseteq {\cal I}_i$ implies $\mathcal{I}_i = \textit{Range} (\mathbf{H}_{ij})$. Therefore, we can assign a Boolean variable $x_{ij}$ to each rank-1 channel $\mathbf{H}_{ij}$,  with ``$x_{ij} = 1$" representing
$\mathcal{S}_j = \textit{Null} (\mathbf{H}_{ij})$ and ``$x_{ij} = 0$" signifying $\mathcal{I}_i = \textit{Range} (\mathbf{H}_{ij})$. In this way, we associate a Boolean variable $x_{ij}$ for each rank-1 crosstalk channel matrix
$\mathbf{H}_{ij}$ in $H$.

Next we represent the interference alignment condition at each receive node of $H$ using the Boolean variables $\{x_{ij}\}$ (plus some auxiliary Boolean variables $\{y_{i},z_{ij},z_{i}\}$ defined below).
Suppose user $i$'s receive node is in $\bH$. We consider the cases $d_i=2$ and $d_i=1$ separately.

{\it Case $d_i=2$}: In this case ${\cal I}_i=0$, so we must have $x_{ij}=1$. We rewrite this condition in the form of two 2-SAT clauses
\begin{equation}\label{eq1}
x_{ij}\vee y_{i},\quad x_{ij}\vee {\bar y}_{i},\quad\mbox{ for all $j\neq i$ and $\rank(\mathbf{H}_{ij}) = 1$},
 \end{equation}
 where $y_{i}$ is an auxiliary Boolean variable. In this case, the satisfaction of \eqref{eq1} and the condition that the receive node of user $i$ is not connected to other {\color{black} users' transmit} nodes via rank-2 links is equivalent to achieving one DoF for user $i$.

{\it {\color{black}Case $d_i=1$} and ${\rm rank}(\mathbf{H}_{ii})=1$}:
In this case, then the received signal subspace is $\mathbf{H}_{ii}{\cal S}_i=\textit{Range}(\mathbf{H}_{ii})$ and $\dim{\cal I}_i=1$, so that all the interference at the receive node of user $i$ must be aligned in {\color{black}an one-dimensional} subspace that is linearly independent of ${\it Range}(\mathbf{H}_{ii})$. We need to further consider several subcases, depending on if the receive node of user~$i$ is connected {\color{black} to other} transmit nodes via rank-1 or rank-2 links. In particular, if the transmit nodes of users $j$ and $k$ are connected to receive node~$i$ via rank-1 links, then the interference alignment condition requires the satisfaction of the following 2-SAT clauses
\begin{equation}\label{eq2}
\begin{array}{rl}
x_{ij}\vee x_{ik},&  \mbox{for all $j\neq k\neq i$ such that ${\rank }(\mathbf{H}_{ij})={\rank }(\mathbf{H}_{ik})=1$ and $\textit{Range} (\mathbf{H}_{ij})\neq \textit{Range} (\mathbf{H}_{ik})$},\\
x_{ij}\vee z_{ij},\ x_{ij}\vee {\bar z}_{ij},&\mbox{for all $j\neq i$ such that ${\rank }(\mathbf{H}_{ij})=1$ and $\textit{Range} (\mathbf{H}_{ij})= \textit{Range} (\mathbf{H}_{ii})$},
\end{array}
\end{equation}
where $z_{ij}$ is a dummy Boolean variable, and the last condition corresponds to the linear independence requirement of the signal/interference subspaces.
Moreover, if there is a rank-2 link connecting the receive node of user $i$ to the transmit node of user $\ell$, $\ell\neq i$, i.e., $\mathbf{H}_{i\ell}$ is full rank, then the receive node of user $i$ is in $\bH'$. Consequently, the transmit strategy of user $\ell$ has only two possibilities B1 and B2 as outlined above.  For the Case B1 where the transmit node of user $\ell$ can pick one of the two possible beamforming vectors $\bv_\ell^0$, $\bv_\ell^1$, we define a Boolean variable $z_{\ell}$ with ``$z_{\ell}=0$" representing
$\bv_\ell^0$ is chosen, while ``$z_{\ell}=1$" signifying $\bv_\ell^1$ is chosen. Now the interference alignment for user $i$ requires the satisfaction of following 2-SAT clauses
\begin{equation}\label{eq4}
\begin{array}{l}
 z_{\ell}\vee x_{ij},\quad \mbox{for all $j\neq\ell\neq i$ such that ${\rank }(\mathbf{H}_{ij})=1$, $\rank(\mathbf{H}_{i\ell})=2$ and $\mathbf{H}_{i\ell}\bv^0_\ell\not\in \textit{Range} (\mathbf{H}_{ij})$},\\
{\bar z}_{\ell}\vee x_{ij},\quad \mbox{for all $j\neq \ell\neq i$ such that ${\rank }(\mathbf{H}_{ij})=1$, $\rank(\mathbf{H}_{i\ell})=2$ and $\mathbf{H}_{i\ell}\bv^1_\ell\not\in \textit{Range} (\mathbf{H}_{ij})$}.
\end{array}
\end{equation}
If in Case B1 the transmit node of user $\ell$ must pick a unique vector $\bv^0_\ell$, then we must have $z_{\ell}=0$ and $x_{ij}=1$ if $\mathbf{H}_{i\ell}\bv^0_\ell\not\in \textit{Range} (\mathbf{H}_{ij})$, and $z_{\ell}=0$ if $\mathbf{H}_{i\ell}\bv^0_\ell\in \textit{Range} (\mathbf{H}_{ij})$. The latter conditions are equivalent to the satisfaction of the following 2-SAT clauses:
\begin{equation}\label{eq5}
\begin{array}{rl}
{\bar z}_{\ell}\vee x_{ij},\ {\bar z}_{\ell}\vee {\bar x}_{ij},\ z_{\ell}\vee x_{ij},& \mbox{for all $j\neq\ell\neq i$ s.t.\ ${\rank }(\mathbf{H}_{ij})=1$, $\rank(\mathbf{H}_{i\ell})=2$ and $\mathbf{H}_{i\ell}\bv^0_\ell\not\in \textit{Range} (\mathbf{H}_{ij})$},\\
{\bar z}_{\ell}\vee x_{ij},\ {\bar z}_{\ell}\vee {\bar x}_{ij},& \mbox{for all $j\neq\ell\neq i$ s.t.\ ${\rank }(\mathbf{H}_{ij})=1$, $\rank(\mathbf{H}_{i\ell})=2$ and $\mathbf{H}_{i\ell}\bv^0_\ell\in \textit{Range} (\mathbf{H}_{ij})$}.
\end{array}
\end{equation}
To ensure linear independence of the signal and interference subspaces for user $i$, we must make sure the satisfaction of the following 2-SAT clauses
\begin{equation}\label{eq7}
\begin{array}{rl}
{\bar z}_{\ell}\vee y_{i},\ {\bar z}_{\ell}\vee {\bar y}_{i}, & \mbox{for all $\ell\neq i$ s.t.\ $\rank(\mathbf{H}_{i\ell})=2$ and $\mathbf{H}_{i\ell}\bv^1_\ell\in \textit{Range} (\mathbf{H}_{ii})$},\\
{z}_{\ell}\vee y_{i},\ { z}_{\ell}\vee {\bar y}_{i}, & \mbox{for all $\ell\neq i$ s.t.\ $\rank(\mathbf{H}_{i\ell})=2$ and $\mathbf{H}_{i\ell}\bv^0_\ell\in \textit{Range} (\mathbf{H}_{ii})$},
\end{array}
\end{equation}
where $y_{i}$ is a dummy Boolean variable.
Now we consider Case B2. Suppose the receive node of user $i$ lies in a connected component $\bH''$ of $\bH'$. Then, {\color{black}for each pair of receive node of users $i$ and $\ell$} in $\bH''$ ($i\neq \ell$), there exists a (efficiently computable) nonsingular matrix $\bG_{i\ell}$ such that
\[
{\cal I}_i=\bG_{i\ell}{\cal I}_\ell.
\]
To ensure this condition, the following 2-SAT clauses must be satisfied for all transmit nodes $j$ and $k$ in $H''$:
\begin{equation}\label{eq3}
x_{ij}\vee x_{\ell k},\quad \mbox{for all $j\neq i,\,k\neq \ell$ s.t. ${\rm rank}(\mathbf{H}_{\ell k})= {\color{black}\rank} (\mathbf{H}_{ij})=1$, and $\bG_{i\ell}\textit{Range} (\mathbf{H}_{\ell k})\neq \textit{Range} (\mathbf{H}_{ij})$}.
\end{equation}
Furthermore, to make sure that the signal and interference subspaces are linearly independent at the receive node of user $i$, we must have for all transmit node $j$ in $H''$ that the following 2-SAT clauses are satisfied
\begin{equation}\label{eq6}
x_{ij}\vee z_{ij},\ x_{ij}\vee {\bar z}_{ij},\quad \mbox{for all $j\neq i$ s.t.\ $\textit{Range}(\mathbf{H}_{ij})= \textit{Range}(\mathbf{H}_{ii})$}.
\end{equation}
Finally, we notice that the Boolean variables $\{x_{i\ell},z_{i\ell}\}$ all represent the signaling strategies of user $\ell$. We must ensure that these signaling strategies are compatible. In other words, we can not simultaneously have both ${\cal S}_\ell=\textit{Null}(\mathbf{H}_{i\ell})$ and ${\cal S}_\ell=\textit{Null}(\mathbf{H}_{j\ell})$ ($j\neq i$), unless of course the two null spaces are equal. This implies that we should have
\begin{equation}\label{eq8}
{\bar x}_{i\ell}\vee {\bar x}_{j\ell},\quad \mbox{for all $i\neq j\neq \ell$ s.t.\ ${\rm rank}(\mathbf{H}_{i\ell})={\rm rank}(\mathbf{H}_{j\ell})=1,\; \textit{Null}(\mathbf{H}_{i\ell})\neq \textit{Null}(\mathbf{H}_{j\ell})$}.
\end{equation}
Moreover, if the transmit node of user $\ell$ is also in $H''$ and its transmit beamforming vector must be chosen from the set $\{\bv^0_\ell,\bv^1_\ell\}$ (Case B1). Then, by a similar argument, we must also ensure the following compatibility conditions:
\begin{equation}\label{eq9}
\begin{array}{rl}
{\bar x}_{i\ell}\vee {z}_{\ell},& \mbox{for all $i\neq j\neq \ell$ s.t.\ ${\rm rank}(\mathbf{H}_{i\ell})=1,\, {\rm rank}(\mathbf{H}_{j\ell})=2,\; {\color{black}\bv^0_{\ell}\not\in \textit{Null}(\mathbf{H}_{i\ell})}$},\\
{\bar x}_{i\ell}\vee {\bar z}_{\ell},& \mbox{for all $i\neq j\neq \ell$ s.t.\ ${\rm rank}(\mathbf{H}_{i\ell})=1,\, {\rm rank}(\mathbf{H}_{j\ell})=2,\; {\color{black}\bv^1_{\ell}\not\in \textit{Null}(\mathbf{H}_{i\ell})}$}.
\end{array}
\end{equation}
In case of B2 (i.e., $H''$ is a tree or all loop matrices are constant multiples of identity matrix), then the transmit subspace of user $\ell$ (which lies in $H''$) can be chosen continuously (rather than {\color{black}from}
a discrete set $\{\bv^0_\ell,\,\bv^1_\ell\}$). In this case, the compatibility condition \eqref{eq8} is sufficient; there is no additional compatibility condition needed.
%

{\it Case $d_i=1$ and ${\rm rank}(\mathbf{H}_{ii})=2$}: In this case, if the transmit node of user $i$ is connected to a receive node of user $j$ via a rank-1 link,  then $x_{ji}=1$ signifies the use of transmit beamforming subspace of ${\it Null}(\mathbf{H}_{ji})$ for user $i$; else if transmitter $i$ is in $H''$ so that its transmit beamforming direction must be chosen from $\bv^0_{i},\,\bv^1_{i}$, corresponding to $z_{i}=0$ and $1$ respectively (Case B1). [Case B2 corresponds to the continuous selection of beamforming vector for user $i$; no 2-SAT clause is needed in that case.] In the first case, the signal subspace at receive node of user $i$ becomes $\mathbf{H}_{ii}{\it Null}(\mathbf{H}_{ji})$, while in the second case, the signal subspace is $\mathbf{H}_{ii}\bv^0_{i}$, or $\mathbf{H}_{ii}\bv^1_{i}$. We must make sure the signal subspace is linearly independent from the interference subspace of user $i$. This implies that the following 2-SAT clauses must be satisfied:
\begin{equation}\label{eq10}
\begin{array}{rl}
{\bar x}_{ji}\vee {x}_{i\ell},& \mbox{for all $i\neq j\neq \ell$ s.t.\ ${\rm rank}(\mathbf{H}_{i\ell})={\rm rank}(\mathbf{H}_{ji})=1,\; \textit{Range}(\mathbf{H}_{i\ell})= \mathbf{H}_{ii}\textit{Null}(\mathbf{H}_{ji})$},\\
{x}_{i\ell}\vee {z}_{i},& \mbox{for all $i\neq j$ s.t.\ ${\rm rank}(\mathbf{H}_{i\ell})=1,\, i\in H'',\; \mathbf{H}_{ii}\bv^0_{i}\in \textit{Range}(\mathbf{H}_{i\ell})$},\\
{x}_{i\ell}\vee {\bar z}_{i},& \mbox{for all $i\neq j\neq \ell$ s.t.\ ${\rm rank}(\mathbf{H}_{i\ell})=1,\, i\in H'',\; \mathbf{H}_{ii}\bv^1_{i}\in \textit{Range}(\mathbf{H}_{i\ell})$}.
\end{array}
\end{equation}
%

It can be checked that the DoF tuple $(d_1,d_2,\ldots,d_K)$ is achievable if and only if conditions \eqref{eq1}-\eqref{eq10} are satisfied for some binary realizations of Boolean variables $\{x_{ij},y_{i},z_{i},z_{ij}\}$. Moreover, the number of such 2-SAT clauses is polynomial in $K$ (in fact $O(K^4)$).
Hence, we have transformed the DoF feasibility problem in polynomial time to an instance of 2-satisfiability problem. The latter problem is known to be solvable in polynomial time.
\end{proof}

%

\section{Strategies for Linear Transceiver Design}
\label{sec:Proposed_Approach}
In this section, we propose linear transceiver design algorithms for interference channels.
Using linear transceivers introduced in Section
\ref{sec:System_Model}, the estimated data stream at receiver~$k$
is given by
\begin{align}
\small
\hat{\mathbf{s}}_k = \mathbf{U}_k^T \sum_{j=1}^K \mathbf{H}_{kj}
\mathbf{V}_j \mathbf{s}_j + \mathbf{U}_k^T \mathbf{n}_k\nonumber
\end{align}
\normalsize
and the SINR value for the $q$-th data stream of user $k$,
$\gamma_k^q$, is given by
\begin{align}
\small
\gamma_k^q = \frac{ |{\mathbf{u}_k^q}^T
\mathbf{H}_{kk}\mathbf{v}_k^q|^2}{\sigma_k^2
\|\mathbf{u}_k^q\|^2+\sum_{(j,r)\neq(k,q)}|{\mathbf{u}_k^q}^T
\mathbf{H}_{kj} \mathbf{v}_j^r|^2}\nonumber
\end{align}
\normalsize where $\mathbf{u}_k^q$ and $\mathbf{v}_k^q$ denote the
$q$-th column of $\mathbf{U}_k$ and $\mathbf{V}_k$, respectively.
Using a linear MMSE receiver $u_k^q$, we have
\begin{align}
\gamma_k^q ={\mathbf{v}_k^q}^T\mathbf{H}_{kk}^T(\sigma^2\mathbf{I}
+\sum_{(j,r)\neq(k,q)}\mathbf{H}_{kj}\mathbf{v}_j^r{\mathbf{v}_j^r}^T
\mathbf{H}_{kj}^T)^{-1}\mathbf{H}_{kk}\mathbf{v}_k^q.\nonumber
\end{align}
\normalsize
One possible choice of the utility function for the $k$-th user
could be the sum of the SINR values of its data streams, i.e.,
\begin{align}
\xi_k&=\sum_{q}\gamma_k^q\nonumber\\
&=\sum_{q}{\mathbf{v}_k^q}^T\mathbf{H}_{kk}^T(\sigma^2\mathbf{I}
+\sum_{(j,r)\neq(k,q)}\mathbf{H}_{kj}\mathbf{v}_j^r{\mathbf{v}_j^r}^T
\mathbf{H}_{kj}^T)^{-1}\mathbf{H}_{kk}\mathbf{v}_k^q.\nonumber
\end{align}
However, maximizing $\xi_k$ does not lead to the
maximization of the total DoF in high SNR.
Therefore, we need to introduce another utility function in order to
capture more DoF for each user.  First, we define ${\cal{U}}(\gamma)=\frac{\gamma}{1+\gamma}$ as the utility function of the $q$-th data stream of user $k$ and then, we consider ${\cal {U}}_k=\sum_q {\cal{U}}(\gamma^q_k)$ as the utility function of user $k$. Thus, at high SNR, ${\cal{U}}_k$ equals the DoF at receiver $k$, while at low SNR, ${\cal{U}}_k$ equals the sum SINR. Using the rank one update of the matrix inverse term in SINR value, we can rewrite ${\cal{U}}_k$ as
\begin{align}
{\cal{U}}_k=\sum_{q}{\mathbf{v}_k^q}^T\mathbf{H}_{kk}^T(\sigma^2\mathbf{I}+
\sum_{(j,r)}\mathbf{H}_{kj}\mathbf{v}_j^r{\mathbf{v}_j^r}^T
\mathbf{H}_{kj}^T)^{-1}\mathbf{H}_{kk}\mathbf{v}_k^q.\nonumber
\end{align}
The proposed utility function preserves fairness among different data streams of user $k$ and also closely approximates the sum DoF at high SNR.

Directly optimizing linear transceivers
$\mathbf{U}_k$'s and $\mathbf{V}_k$'s requires specification of DoFs $d_k$ in advance, since the
dimension of $\mathbf{U}_k$ and $\mathbf{V}_k$ depends on $d_k$.
To avoid this explicit dependence on $d_k$,  we consider optimizing the
transmit covariance matrix instead of linear transceivers $\mathbf{U}_k$ and $\mathbf{V}_k$. In particular, we write the utility function of user $k$ as
\begin{align}
{\cal{U}}_k=\tr\left[\mathbf{H}_{kk}\mathbf{Q}_{k}\mathbf{H}_{kk}^T(\sigma^2\mathbf{I}+\sum_{\ell=1}^k\mathbf{H}_{k
\ell}\mathbf{Q}_{\ell}\mathbf{H}_{k\ell}^T)^{-1}\right]\label{EQ:U_k}
\end{align}
\normalsize
where $\mathbf{Q}_{k}=\sum_q \mathbf{v}_k^q(\mathbf{v}_k^q)^T$ is
the transmit covariance matrix of the $k$-th user. However, this utility function still does not related to the sum-rate directly. In the sequel, we propose a weighting approach to relate the utility function in \eqref{EQ:U_k} to the rate of user~$k$.

Consider the well-known weighted sum-rate maximization problem
\begin{align}
& \max_{\{\mathbf{Q}_k\}_{k=1}^K}\quad \sum_{k=1}^K \alpha_k R_k \label{EQ:WSRM_Original}\\
& \st \;\; \tr(\mathbf{Q}_{k}) \leq p_{k},\quad \quad\mathbf{Q}_{k}\succcurlyeq 0\nonumber
\end{align}
\normalsize
where  $R_k \triangleq \log \det \left(\mathbf{I} + \mathbf{H}_{kk} \mathbf{Q}_k \mathbf{H}_{kk}^T (\sigma^2\mathbf{I} +\sum_{\ell \neq k}\mathbf{H}_{k \ell}\mathbf{Q}_{\ell} \mathbf{H}_{k\ell}^T)^{-1}\right) $ \normalsize  is the achievable rate of user~$k$ and the coefficient  $\alpha_k$ denotes user~$k$'s weight. Using linear algebra to simplify the objective function, the above problem can be reformulated as
the following equivalent optimization problem:
\begin{align}\label{EQ:Weighted_Sum_Rate}
&\min_{\{\mathbf{Q}_k\}_{k=1}^K}  \quad \sum_{k=1}^K  \alpha_k \log \det \left( \mathbf{I} - \mathbf{H}_{kk} \mathbf{Q}_k \mathbf{H}_{kk}^T \left(\sigma^2\mathbf{I}+\sum_{\ell=1}^K\mathbf{H}_{k
\ell}\mathbf{Q}_{\ell}\mathbf{H}_{k\ell}^T\right)^{-1}\right) \\
& \st \quad \tr(\mathbf{Q}_{k}) \leq p_{k},\quad \quad\mathbf{Q}_{k}\succcurlyeq 0,\nonumber
\end{align}
\normalsize
where the term inside the determinant is linearly related to the utility function in \eqref{EQ:U_k}. {\color{black} Similar to \cite{cioffi-BC}, we reformulate the problem \eqref{EQ:Weighted_Sum_Rate} by further introducing new optimization variables $\mathbf{W}_k \in \mathbb{R}^{M \times M},\ k=1,2,\ldots, K,$ to obtain the following equivalent optimization problem}
\begin{align}
&\min_{\{\mathbf{Q}_k , \mathbf{W}_k\}_{k=1}^K} \quad \sum_{k=1}^K \alpha_k \tr (\mathbf{W}_k g_k(\mathbf{Q})) - \sum_{k=1}^K \alpha_k \log \det \mathbf{W}_k\label{EQ:equivalent_problem}\\
& \st \quad \tr(\mathbf{Q}_{k}) \leq p_{k}\quad \quad\mathbf{Q}_{k}\succcurlyeq 0,\nonumber
\end{align}
\normalsize
{\color{black} where $\mbQ = (\mbQ_1, \mbQ_2,\ldots,\mbQ_K)$ and}
\begin{align}
g_k(\mathbf{Q}) \triangleq \mathbf{I} -  \mathbf{H}_{kk} \mathbf{Q}_k \mathbf{H}_{kk}^T (\sigma^2\mathbf{I}+\sum_{\ell=1}^K\mathbf{H}_{k
\ell} \mathbf{Q}_{\ell} \mathbf{H}_{k\ell}^T)^{-1}.\nonumber
\end{align}
\normalsize The optimization problem \eqref{EQ:equivalent_problem} is convex in $\{\mathbf{W}_k\}_{k=1}^K$. By checking the first order optimality condition,  the optimal  $\mathbf{W}_k$ is given by
\begin{align}
\mathbf{W}_k^{opt} =   \mathbf{I} + \mathbf{H}_{kk} \mathbf{Q}_k \mathbf{H}_{kk}^T (\sigma^2\mathbf{I}+\sum_{\ell \neq k} \mathbf{H}_{k \ell}\mathbf{Q}_{\ell}\mathbf{H}_{k\ell}^T)^{-1}, \quad \forall \ k = 1,2,\ldots, K. \label{EQ:Weight_Update}
\end{align}
\normalsize
By plugging back the optimal $\mathbf{W}_k^{opt}$ in \eqref{EQ:equivalent_problem}, we immediately see the equivalence of \eqref{EQ:equivalent_problem} and \eqref{EQ:Weighted_Sum_Rate}. Furthermore, in order to have a distributed approach, we let users update their transmit covariance matrix independently. Therefore, for fixed  $\{\mathbf{W}_k\}_{k=1}^K$, user~$k$ can solve the following optimization problem to update its transmit covariance matrix:
\begin{align}
& \max_{\mathbf{Q}_k} \quad \alpha_k \tr
\left[\mathbf{W}_k \mathbf{H}_{kk}\mathbf{Q}_{k}\mathbf{H}_{kk}^T(\sigma^2\mathbf{I}+
\sum_{l=1}^K\mathbf{H}_{k\ell}\mathbf{Q}_{\ell}\mathbf{H}_{k\ell}^T)^{-1}\right] \nonumber\\
&\quad \quad+ \sum_{j \neq k} \alpha_j \tr \left[\mathbf{W}_j \mathbf{H}_{jj}\mathbf{Q}_{j}\mathbf{H}_{jj}^T(\sigma^2\mathbf{I}+ \sum_{l=1}^K\mathbf{H}_{j\ell}\mathbf{Q}_{\ell}\mathbf{H}_{j\ell}^T)^{-1}\right] \label{EQ:Original_Obj}\\
& \st \;\;\; \tr(\mathbf{Q}_{k}) \leq p_{k}\;\;\; \;\;\;\; \mathbf{Q}_{k}\succcurlyeq 0.\nonumber
\end{align}
\normalsize
Unfortunately, this objective function is not convex. In order to make the
problem convex, we keep the first term in the objective function
(which is a concave function of $\mathbf{Q}_k$) and use the
local linear approximation of the second term, i.e.,
\begin{eqnarray*}
 \sum_{j \neq k} \alpha_j \; && \!\!\!\!\!\!\!\!\!\!\!\!\!\!\!\!\tr \left[\mathbf{W}_j \mathbf{H}_{jj}\mathbf{Q}_{j}\mathbf{H}_{jj}^T(\mathbf{C}_{jk} + \mathbf{H}_{jk}\mathbf{Q}_k\mathbf{H}_{jk}^T)^{-1}\right]\nonumber\\
&\approx & \sum_{j \neq k} \alpha_j \tr
\left\{\mathbf{W}_j \mathbf{H}_{jj}\mathbf{Q}_{j}\mathbf{H}_{jj}^T
\left[(\mathbf{C}_{jk}+
\mathbf{H}_{jk}\widetilde{\mathbf{Q}}_k\mathbf{H}_{jk}^T)^{-1}\right.\right.\nonumber\\
&&- (\mathbf{C}_{jk}+
\mathbf{H}_{jk}\widetilde{\mathbf{Q}}_k\mathbf{H}_{jk}^T)^{-1}
\mathbf{H}_{jk} \mathbf{Q}_k \mathbf{H}_{jk}^T (\mathbf{C}_{jk}+
\mathbf{H}_{jk}\widetilde{\mathbf{Q}}_k\mathbf{H}_{jk}^T)^{-1} \nonumber\\
&&+\left. \left.(\mathbf{C}_{jk}+
\mathbf{H}_{jk}\widetilde{\mathbf{Q}}_k\mathbf{H}_{jk}^T)^{-1}
\mathbf{H}_{jk} \widetilde{\mathbf{Q}}_k \mathbf{H}_{jk}^T
(\mathbf{C}_{jk}+
\mathbf{H}_{jk}\widetilde{\mathbf{Q}}_k\mathbf{H}_{jk}^T)^{-1}\right]\right\}\nonumber
\end{eqnarray*}
\normalsize where $\widetilde{\mathbf{Q}}_k$ is the local value of
transmit covariance matrix at the previous iteration and
$\mathbf{C}_{jk}$ is the received signal covariance matrix at receiver~$j$ excluding the $k$-th user's signal, i.e.,
\begin{align}
\mathbf{C}_{jk} \triangleq \sigma^2 \mathbf{I} + \sum_{\ell \neq k}
\mathbf{H}_{j\ell} \mathbf{Q}_\ell \mathbf{H}_{j\ell}^T. \label{eq:C}
\end{align}
By substituting the above approximation in \eqref{EQ:Original_Obj} and simplifying the resulting optimization problem,
we get
\begin{equation}\label{EQ:Concave_Lower_Bound}
\begin{array}{rl}
\displaystyle
 \max_{\mathbf{Q}_k} & \displaystyle \alpha_k \tr
\left[\mathbf{W}_k \mathbf{H}_{kk}\mathbf{Q}_{k}\mathbf{H}_{kk}^T(\sigma^2\mathbf{I}+
\sum_{l=1}^K\mathbf{H}_{k\ell}\mathbf{Q}_{\ell}\mathbf{H}_{k\ell}^T)^{-1}\right] -
 \tr \left[ \mathbf{B}_k \mathbf{Q}_k \right] \\
\mbox{s.t. } & \tr(\mathbf{Q}_{k}) \leq p_{k}\;\;\;\; \;\;\;\;
\mathbf{Q}_{k}\succcurlyeq 0
\end{array}
\end{equation}
\normalsize where \small $\mathbf{B}_k \triangleq \sum_{j \neq k}
\mathbf{H}_{jk}^T (\mathbf{C}_{jk}+
\mathbf{H}_{jk}\widetilde{\mathbf{Q}}_k\mathbf{H}_{jk}^T)^{-1}
\alpha_j \mathbf{W}_j \mathbf{H}_{jj}\mathbf{Q}_j \mathbf{H}_{jj}^T (\mathbf{C}_{jk}+
\mathbf{H}_{jk}\widetilde{\mathbf{Q}}_k\mathbf{H}_{jk}^T)^{-1}
\mathbf{H}_{jk}$. \normalsize
{\color{black} The objective function in \eqref{EQ:Concave_Lower_Bound} considers the effect of transmit covariance matrix of user~$k$ on not only its own rate, but also those of others in the interference channel. Similar balanced approaches have been considered in related works, see \cite{ILA,gesbert1,gesbert2,gesbert3}.} By further simplification of the objective function and using the Schur complement, the problem can be formulated as the following Semi-definite Programming (SDP) form:
\begin{align}
  & \min_{\mathbf{Q}_k ,\mathbf{Y}} \;\; \alpha_k \tr \left[\mathbf{Y}\right] +  \tr \left[ \mathbf{B}_k \mathbf{Q}_k \right] \label{EQ:SDP} \\
    & \st \;\;\; \tr(\mathbf{Q}_{k}) \leq p_{k} ,\;\;\;\;\mathbf{Q}_{k}\succcurlyeq 0,\nonumber\\
    &  \left[
         \begin{array}{cc}
           \mathbf{C}_{kk}+ \mathbf{H}_{kk} \mathbf{Q}_k \mathbf{H}_{kk}^T  &(\mathbf{W}_{k}\mathbf{C}_{kk})^{1/2}  \\
            (\mathbf{W}_{k}\mathbf{C}_{kk})^{1/2} & \mathbf{Y} \\
          \end{array}
\right] \succcurlyeq 0. \nonumber
\end{align}
Note that the matrices $\mathbf{W}_k$ and $\mathbf{C}_{kk}$ are updated by \eqref{EQ:Weight_Update} and \eqref{eq:C} respectively. Thus $\mathbf{W}_k \mathbf{C}_{kk}$ is Hermitian positive semi-definite. Hence, for fixed matrices $\{\mathbf{W}_k\}_{k=1}^K$, user~$k$ can update its
transmit covariance matrix $\mathbf{Q}_k$ by solving the above SDP
problem. 
\begin{figure}[htbp]
\centering
\begin{tabular}{|p{4.3in}|}
\hline
~\,1.\quad initialize with $\mathbf{Q}_k = \frac{p_k}{M_k} \mathbf{I}$ and  $\mathbf{W}_k = \mathbf{I}$, for all $k = 1,2,\cdots,K$\\
~\,2.\quad \textbf{repeat}\\
~\,3.\quad \quad \textbf{for} $k = 1,2,\cdots,K$ \textbf{do}\\
~\,4.\quad \quad \quad update $\mathbf{W}_k$ according to \eqref{EQ:Weight_Update}\\
~\,5.\quad \quad \quad update $\mathbf{Q}_k$ by solving \eqref{EQ:SDP}\\
~\,6.\quad \quad \quad update $\mathbf{W}_k$ according to \eqref{EQ:Weight_Update}\\
~\,7.\quad \textbf{until} convergence, or $\|\mathbf{\tilde{Q}} - \mathbf{Q}\| \leq \epsilon$\\
\hline
\end{tabular}\vspace{1.2em}
\caption{An iterative SDP approximation algorithm for sum-rate maximization} \label{fig:code}
\end{figure}
%

Note that the second term in \eqref{EQ:Original_Obj} is a convex
function of $\mathbf{Q}_k$. Therefore, the local linear
approximation is a lower bound which is tight at the current point
$\widetilde{\mathbf{Q}}_k$. Hence, by solving \eqref{EQ:SDP}, we
minimize a concave lower bound of the original utility function
\eqref{EQ:Original_Obj}. Since the previous iterate $\widetilde
{\mathbf Q}_k$ is feasible for \eqref{EQ:Original_Obj}, it
follows that the system utility function (i.e., the objective
function of \eqref{EQ:Original_Obj}) is non-decreasing.
Furthermore, \eqref{EQ:Original_Obj} is bounded from above and
this implies the sequence of objective function values generated by
the proposed algorithm converges. The following theorem further
establishes the iterate convergence to a stationary point for the
proposed algorithm.
In order to prove that each limit point of this algorithm is a stationary point of the original problem we need the following lemma.\\
\begin{lemma}
\label{lemma: str_concavity}
 If the direct  channel matrices are full-rank and tall, the function:
\begin{align}
\bar{f}_{2k-1}(\mathbf{Q}_k) \triangleq \alpha_k \tr
\left[\mathbf{W}_k \mathbf{H}_{kk}\mathbf{Q}_{k}\mathbf{H}_{kk}^T(\sigma^2\mathbf{I}+
\sum_{l=1}^K\mathbf{H}_{k\ell}\mathbf{Q}_{\ell}\mathbf{H}_{k\ell}^T)^{-1}\right] -
 \tr \left[ \mathbf{B}_k \mathbf{Q}_k \right]
\end{align}
is strictly concave with respect to symmetric positive semidefinite matrix $\mathbf{Q}_k$. Moreover, the objective function of \eqref{EQ:equivalent_problem} is also strictly convex with respect to $\mathbf{W}_k$.
\end{lemma}
\begin{proof}
See Appendix, Section \ref{sec:appendix}.
\end{proof}
\begin{thm}
\label{thm: convergence}  Assuming that the direct channel matrices, $\mathbf{H}_{kk}$, are full rank and tall, then every limit point
 of the proposed algorithm is a stationary point of \eqref{EQ:WSRM_Original}.
\end{thm}
\begin{proof}
According to Lemma \ref{lemma:Stationarity} (see Appendix, section \ref{sec:appendix}), every stationary point of \eqref{EQ:equivalent_problem} is also a stationary point of \eqref{EQ:WSRM_Original}. Therefore, we only need to prove that every limit point of the proposed algorithm is a stationary point of \eqref{EQ:equivalent_problem}. To this end, let us define the auxiliary variable $\mathbf{X}^i = \{\mathbf{X}_{\ell}^i\}_{\ell = 1}^{2K}$, where $\mathbf{X}_{2k-1}^i \triangleq \mathbf{Q}_k^i$ is the updated transmit covariance matrix of user~$k$ at $i$-th iteration and $\mathbf{X}_{2k}^i \triangleq \mathbf{W}_k^i$ is the updated weight matrix of user~$k$ at $i$-th iteration. In particular, we define $\mathbf{Q}_k^i$ to be the solution of the following problem
\begin{align}
\mathbf{X}_{2k-1}^i = \mathbf{Q}_k^i\triangleq \; &\text{Arg} \displaystyle{\max_{\mathbf{Q}_k}} \quad \bar{f}_{2k-1} (\mathbf{Q}_k; \mathbf{X}_1^i, \mathbf{X}_2^i, \ldots, \mathbf{X}_{2k-2}^i, \mathbf{X}_{2k-1}^{i-1}, \mathbf{X}_{2k}^{i-1},\ldots , \mathbf{X}_{2K}^{i-1}) \nonumber\\
& \st \quad \tr(\mathbf{Q}_{k}) \leq p_{k},\quad\quad
\mathbf{Q}_{k}\succcurlyeq 0 \nonumber
\end{align}
where $\bar{f}_{2k-1} (\mathbf{Q}_k; \mathbf{X}_1^i, \mathbf{X}_2^i, \ldots, \mathbf{X}_{2k-2}^i, \mathbf{X}_{2k-1}^{i-1}, \mathbf{X}_{2k}^{i-1},\ldots , \mathbf{X}_{2K}^{i-1})$ is the objective function of \eqref{EQ:Concave_Lower_Bound} which is the local concave lower bound approximation of the objective function in \eqref{EQ:equivalent_problem} as discussed in section \ref{sec:Proposed_Approach}. Similarly, we define $\mathbf{X}_{2k}^i = \mathbf{W}_k^i$ to be the updated weight matrix of user~$k$ at iteration~$i$, i.e.,
\begin{align}
\mathbf{X}_{2k}^i = \mathbf{W}_k^i\triangleq \; &\text{Arg} \displaystyle{\max_{\mathbf{W}_k}} \quad {f} (\mathbf{W}_k; \mathbf{X}_1^i, \mathbf{X}_2^i, \ldots, \mathbf{X}_{2k-1}^i, \mathbf{X}_{2k+1}^{i-1}, \mathbf{X}_{2k}^{i-1},\ldots , \mathbf{X}_{2K}^{i-1}) \nonumber
\end{align}
where $f(\cdot;\cdot)$ is the objective function in \eqref{EQ:equivalent_problem}. \\

Let $\mathbf{X}^i \triangleq (\mathbf{X}_1^i, \mathbf{X}_2^i, \ldots, \mathbf{X}_{2K}^i)$ be the tuple of transmit covariance--weight matrices and $\mathbf{X}^*$ be a limit point of  the sequence $\{\mathbf{X}^i\}_{i=1}^{\infty}$. Therefore, there exists a subsequence of indices $\{i_1,i_2,...,i_j,...\}$ such that
\begin{align}
\lim_{j\rightarrow \infty} \mathbf{X}^{i_j} = \mathbf{X}^* \nonumber
\end{align}
First, we will prove that $\displaystyle{\lim_{j \rightarrow \infty}} \mathbf{X}_1^{i_j+1} - \mathbf{X}_1^{i_j} = \mathbf{0}$ by using contradiction. Suppose the contrary. Hence, by further restricting to a subsequence if necessary, we have
\begin{align}
\exists \gamma^*>0 \;\;\text{such that}\;\; \gamma^{i_j} \geq \gamma^*,\quad \;\;\forall j,\nonumber
\end{align}
where $\gamma^{i_j} = \| \mathbf{X}_1^{i_j+1} - \mathbf{X}_1^{i_j} \|$. Let $\mathbf{S}_1^{i_j} \triangleq \frac{\mathbf{X}_1^{i_j+1}-\mathbf{X}_1^{i_j}}{\gamma^{i_j}}$. Since $\| \mathbf{S}_1^{i_j}\| = 1$, according to Bolzano-Weierstrass theorem, there exists a subset of indices, denoted by $I$, and a unit length matrix $\mathbf{S}_1^*$ such that
\begin{align}
\displaystyle{\lim_{i_j\in I,\; j \to \infty}} \mathbf{S}_1^{i_j} = \mathbf{S}_1^*. \nonumber
\end{align}
Obviously, $0 \leq \epsilon \gamma^* \leq \gamma^{i_j}$  for every $\epsilon$, $0 \leq \epsilon \leq 1$. Moreover, since the feasible set is convex, $\mathbf{X}_1^{i_j} + \epsilon \gamma^* \mathbf{S}_1^{i_j}$ belongs to the feasible set. Therefore, according to the definition of $\mathbf{X}_1^{i_j+1}$ and using the concavity of $\bar{f}_1$, we have
\begin{align}
\bar{f}_1 (\mathbf{X}_1^{i_j+1}; \mathbf{X}^{i_j})
\geq
\bar{f}_1 (\mathbf{X}_1^{i_j} + \epsilon \gamma^* \mathbf{S}_1^{i_j}; \mathbf{X}^{i_j})
\geq
\bar{f}_1 (\mathbf{X}_1^{i_j}; \mathbf{X}^{i_j}). \label{EQ: Sandwich}
\end{align}
On the other hand, the value of the objective function in \eqref{EQ:equivalent_problem} is always increasing and  bounded from above. Moreover, the feasible set is closed and therefore $\mathbf{X}^*$ is in the feasible set. Hence, the value of objective function converges to $f(\mathbf{X}^*)$, i.e.,
\begin{align}
\lim_{j \rightarrow \infty} \bar{f}_1 (\mathbf{X}_1^{i_j}; \mathbf{X}^{i_j}) =
\lim_{j \rightarrow \infty} \bar{f}_1 (\mathbf{X}_1^{i_j+1}; \mathbf{X}^{i_j}) =
f(\mathbf{X^*}). \nonumber
\end{align}
Therefore, letting $j\to \infty$ with ${i_j\in I}$ in \eqref{EQ: Sandwich} yields
\begin{align}
\bar{f}_1 (\mathbf{X}_1^{*} + \epsilon \gamma^* \mathbf{S}_1^*; \mathbf{X}^{*}) =
f(\mathbf{X^*}),\quad \forall\ \epsilon \in [0, 1], \nonumber
\end{align}
which contradicts the strict concavity of $\bar{f}_1(\cdot)$ (c.f.\ Lemma \ref{lemma: str_concavity}). {\color{black} Therefore,  $\displaystyle{\lim_{j \rightarrow \infty}} \mathbf{X}_1^{i_j+1} - \mathbf{X}_1^{i_j} = \mathbf{0}$, or equivalently, we have
\begin{align}
\lim_{j \to \infty} \mathbf{X}_1^{i_j+1} = \lim_{j \to \infty} \mathbf{X}_1^{i_j} = \mathbf{X}_1^*. \label{eq:limit_exists}
\end{align}
On the other hand, $\mathbf{X}_1^{i_j+1}$ is the local maximum of  $\bar{f}_1(\cdot, \mathbf{X}^{i_j})$. Hence,
\begin{align}
\tr \left[{\nabla_{\mathbf{X}_1} \bar{f}_1 (\mathbf{X}_1^{i_j + 1}; \mathbf{X}^{i_j})}^T (\mathbf{X}_1 - \mathbf{X}_1^{i_j})\right] \leq 0, \nonumber
\end{align}
for any feasible point $\mathbf{X}_1$.
Letting $j \to \infty$ and using \eqref{eq:limit_exists} yield
\begin{align}
\tr \left[{\nabla_{\mathbf{X}_1} \bar{f}_1 (\mathbf{X}_1^{*}; \mathbf{X}^{*})}^T (\mathbf{X}_1 - \mathbf{X}_1^{*})\right] \leq 0. \nonumber
\end{align}
Since $f(\cdot)$ and $\bar{f}_1(\cdot, \mathbf{X}^{*})$ have the same gradient with respect to $\mathbf{X}_1^{*}$ at point $\mathbf{X}^{*}$, it follows that}
\begin{align}
\tr \left[{\nabla_{\mathbf{X}_1}  f(\mathbf{X}^*)}^T (\mathbf{X}_1 - \mathbf{X}_1^*)\right] \leq 0. \nonumber
\end{align}
Repeating the same argument for all $k=1,2, \ldots, 2K$, we get
\begin{align}
\tr \left[{\nabla_{\mathbf{X}_k} f(\mathbf{X}^*)}^T (\mathbf{X}_k - \mathbf{X}_k^*)\right] \leq 0,\quad \forall\ k = 1,2,\ldots,2K. \nonumber
\end{align}
By summing up all the equations for all $k$'s we get,
\begin{align}
\tr \left[{\nabla_{\mathbf{X}} f(\mathbf{X}^*)}^T (\mathbf{X} - \mathbf{X}^*)\right] \leq 0 \nonumber
\end{align}
which implies the stationarity of  $\mathbf{X}^*$.
\end{proof}

A couple of remarks are in order. First, in the proof of Theorem~\ref{thm: convergence} we have only used the strict concavity of function $\bar{f}(\cdot)$. Consequently, the proof works for other  objective functions that have the same property and using similar methods, e.g. \cite{ILA}.
Second, after solving \eqref{EQ:SDP} to get the
solution $\mathbf{Q}_k^*$, we can update the transmit covariance
matrix by using relaxation parameter $0 < \alpha \leq 1$, i.e.,
\small$\mathbf{Q}_k \longleftarrow \alpha \mathbf{Q}_k^* +
(1-\alpha) \widetilde{\mathbf{Q}}_k$.\normalsize\; It can be shown
that the convergence result of Theorem \ref{thm: convergence}
holds even by using a fixed relaxation parameter.

An alternative to solving \eqref{EQ:SDP} at each
iteration is to update the transmit covariance matrix in a
totally unselfish manner, i.e., solving the following problem
\begin{align}
  & \min_{\mathbf{Q}_k } \;\;  \tr \left[ \mathbf{B}_k \mathbf{Q}_k \right] \nonumber\\
    & \st \;\;\; \tr(\mathbf{Q}_{k}) = p_{k} ,\;\;\;\;\mathbf{Q}_{k}\succcurlyeq 0.
\end{align}
The above problem has a closed form solution $\mathbf{Q}_k = p_k \mathbf{q}\mathbf{q}^T$, where $\mathbf{q}$ is the eigen vector of $\mathbf{B}_k$ corresponding to its minimum eigen value. This unselfish approach requires all the users to exhaust all their
transmit power, potentially causing unnecessary interference. Furthermore, it results in one DoF for each user because $\mathbf{Q}_k$ is always rank one. In cases that the all one DoF vector is not appropriate either because it is not achievable or because it is too conservative, the above unselfish strategy cannot lead to the maximization of
sum DoFs.

In general, if we know the DoF of each user a-priori and allocate equal power across the data streams,
we can update the transmit beamformer of user~$k$ by solving the following optimization problem:
\begin{align}
  & \min_{\mathbf{V}_k } \;\;  \tr \left[ \mathbf{V}_k^T \mathbf{B}_k \mathbf{V}_k \right] \label{EQ:Unselfish}\\
    & \st \;\;\; \mathbf{V}_k^T \mathbf{V}_k  =\frac{p_k}{d_k}\;\; \mathbf{I}. \nonumber
\end{align}
This approach lets each transmitter use maximum power and pick a transmit covariance matrix $\mathbf{V}_k$ so as to minimize the total interference to other users.  It has a closed form solution $\mathbf{V}_k$ whose columns are proportional to the eigenvectors of $\mathbf{B}_k$ corresponding to its $d_k$ smallest eigen values, scaled appropriately to satisfy the power budget constraint. 
\begin{figure}[htbp]
\centering
\begin{tabular}{|p{4.3in}|}
\hline
~\,1.\quad initialize with $\mathbf{V}_k = \mathbf{0}$ and  $\mathbf{W}_k = \mathbf{I}$, for all $k = 1,2,\cdots,K$\\
~\,2.\quad \textbf{repeat}\\
~\,3.\quad \quad \textbf{for} $k = 1,2,\cdots,K$ \textbf{do}\\
~\,4.\quad \quad \quad update $\mathbf{V}_k$ by solving \eqref{EQ:Unselfish}\\
~\,5.\quad \quad \quad update $\mathbf{W}_k$ according to \eqref{EQ:Weight_Update}\\
~\,6.\quad \textbf{until} convergence\\
\hline
\end{tabular}\vspace{1.2em}
\caption{The unselfish algorithm for sum DoF maximization} \label{fig:code_unselfish}
\end{figure}

\section{Simulation Results}
\label{sec:Simulations} In this section, we present some numerical
results comparing the Decentralized Interference Alignment (DIA)
method \cite{Jafar2} with our proposed methods. All numerical
results are averaged over 20 channel realizations. In each channel
realization, the path loss of the channel coefficients are generated by a relay-backhaul model provided by Huawei Technologies. We consider 19-hexagonal wrap-around cell layout. We randomly choose $K$ base stations, each serving a random relay in its own cell at each time slot. Each base station serves different relays in its own cell orthogonally. Therefore, at each time slot, the base station-relays form an interference channel. The relays have fixed locations so the the system has enough time to learn the channels. The MIMO channel coefficients are modeled by the standard single tap Rayleigh fading model.
We consider linear MMSE receivers and equal power budget for all
users and for all methods.
To implement DIA, we need to predetermine DoF for all users. In
all simulations DoFs are set to be equal for all users.

\begin{figure}[ht!]
 \centering
\includegraphics[width=9cm]{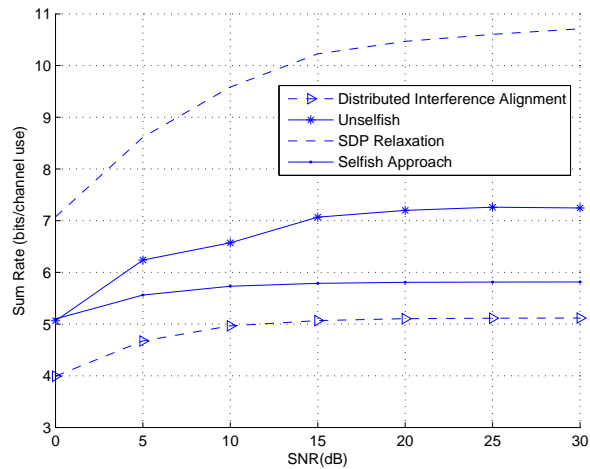}
\caption{Sum-rate vs. SNR: $K=10, M=2, d=1$}\label{FIG:K=10}
\end{figure}

In the first numerical experiment, we consider $K=10$ base station-relay pairs, each
equipped with $M=2$ antennas. The predetermined degrees of freedom
used in DIA method are $d_1=d_2=\ldots=d_K=d=1$. Figure
\ref{FIG:K=10} represents the sum-rate comparison between the
proposed methods and DIA. As Figure \ref{FIG:K=10} shows, the
proposed method yields substantially higher sum-rates in this case.
In fact, the sum-rate achieved by the DIA method
does not grow linearly with SNR, indicating that interference
alignment has not been achieved.

\begin{figure}[ht!]
\centering
\includegraphics[width=9cm]{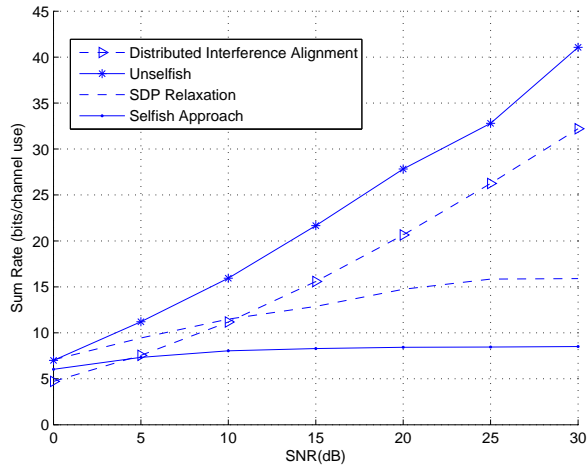}
\caption{Sum-rate vs. SNR: $K=4, M=3, d=1$}\label{FIG:K=4}
\end{figure}

It is known that the DIA method
works well for the $K=3$ case where interference alignment is possible \cite{Jafar2}. We consider the case of $K=3$ transceiver pairs each equipped with $M=3$ antennas and one DoF is considered for each transmitter. As can be seen in Fig. \ref{FIG:K=4}, the selfish and the SDP approach works well in low SNR, but is outperformed by the DIA approach in high SNR region where the interference alignment effect begins to kick in. Interestingly, our Unselfish approach for interference alignment outperforms the DIA algorithm in the entire practical SNR range. Although the DIA method and the Unselfish approach both achieve a sum-rate that increases linearly with SNR, the Unselfish approach has a better offset compared to the DIA method.

\section{Appendix: Proof of Lemmas \ref{lemma: str_concavity} and \ref{lemma:Stationarity}}
\label{sec:appendix}
\setcounter{lemma}{0}
\begin{lemma}
\label{lemma: str_concavity}
 If the direct  channel matrices are full-rank and tall, the function:
\begin{align}
\bar{f}_{2k-1}(\mathbf{Q}_k) \triangleq \alpha_k \tr
\left[\mathbf{W}_k \mathbf{H}_{kk}\mathbf{Q}_{k}\mathbf{H}_{kk}^T(\sigma^2\mathbf{I}+
\sum_{l=1}^K\mathbf{H}_{k\ell}\mathbf{Q}_{\ell}\mathbf{H}_{k\ell}^T)^{-1}\right] -
 \tr \left[ \mathbf{B}_k \mathbf{Q}_k \right]
\end{align}
is strictly concave with respect to symmetric positive semidefinite matrix $\mathbf{Q}_k$. Moreover, the objective function of \eqref{EQ:equivalent_problem} is also strictly convex with respect to $\mathbf{W}_k$.
\end{lemma}
\begin{proof}
Using the notations we have
 defined so far, $\bar f_{2k-1}(\mathbf{Q}_k)$ is given by
\begin{align}
\bar{f}_{2k-1}(\mathbf{Q}_k) = \alpha_k  \left[ \tr\left(\mathbf{W}_k\mathbf{C}_{kk}(\mathbf{C}_{kk}+\mathbf{H}_{kk}\mathbf{Q}_{k}\mathbf{H}_{kk}^{T})^{-1}\right)-\tr{(\mathbf{B}_k\mathbf{Q}_k)}\right].\nonumber
\end{align}
 The second term in $\bar{f}_{2k-1}(\mathbf{Q}_k)$ is linear in $\mathbf{Q}_k$ and does not change the strict concavity of the function.
Hence, it suffices to show the strict concavity of $\tr\left(\mathbf{W}_k\mathbf{C}_{kk}(\mathbf{C}_{kk}+\mathbf{H}_{kk}\mathbf{Q}_{k}\mathbf{H}_{kk}^{T})^{-1}\right)$.  To do so, it is enough to prove that the function is strictly concave in any feasible direction. We drop the index $k$ 
for notational simplicity.
Let us consider a feasible direction denoted by a symmetric matrix $\mathbf{D}\neq 0$ of appropriate size  and a scalar $t\ge0$. We further define the notation $\mathbf{D}^{'}=\mathbf{HD}\mathbf{H}^{T}$ and the function
\begin{align}
h_1(t)=\tr{(\mathbf{WC}(\mathbf{C+X}+t\mathbf{D}^{'}))}\nonumber
\end{align}
where $\mathbf{X=HQ}\mathbf{H}^{T}$ and $\mathbf{C}+\mathbf{X}+t\mathbf{D}^{'}$ are positive definite matrices. Since the direct channel matrix $\bf H$ is tall and full-rank, it follows that $\mathbf{D}^{'}\neq \mathbf{0}$. Moreover,
by the definitions of $\bW$ and $\bC$ \eqref{EQ:Weight_Update}-\eqref{eq:C}, we know the matrix $\mathbf{W}\mathbf{C}$ is symmetric and positive definite.
It suffices to show the strict concavity of $h_1$ with respect to $t$ for each symmetric $\mathbf{D}$.

To prove the strict concavity of $h_1$,
we will calculate the second order derivative of $h_1$ with respect to $t$ and prove that it is negative.
If we denote $\mathbf{B}=(\mathbf{C}+\mathbf{X}+t\mathbf{D}^{'})^{-1}$, then the first order derivative is given by
\begin{align}
\frac{\partial h_1}{\partial t}=\frac{\partial \tr \left(
\mathbf{WC(C+X}+t\mathbf{D^{'}})^{-1}\right)}{\partial t}=\tr \left(
\left(\nabla_{\mathbf{B}} \tr(\mathbf{WCB})\right)^{T}\frac{\partial
\mathbf{B}}{\partial t}\right).\nonumber
\end{align}
In addition, we know that
\begin{align}
\frac{\partial \mathbf{B}}{\partial t}=-\mathbf{B}\mathbf{D^{'}}\mathbf{B}\nonumber,\\
 {\nabla_{\mathbf{B}} \tr\mathbf{(WCB)}}=\mathbf{C}^{T}\mathbf{W}^{T}\nonumber,
\end{align}
which further implies
\begin{align}
\frac {\partial h_1}{\partial t}=-\tr\left(\mathbf{WCBD^{'}B}\right).
\end{align}
In a similar way, we can calculate the second order derivative $\frac {\partial^2 h_1}{\partial t^2}$
\begin{align}
\frac {\partial^2h_1}{{\partial
t}^2}=-2\tr \left(\mathbf{D^{'}BWCBD^{'}B}\right)=-2\tr\left(\mathbf{(D^{'}BD^{'})(BWCB)}\right).
\end{align}
As $\mathbf C$ and $\mathbf B$ are positive definite we can conclude
that $\mathbf{BWCB}$ is also positive definite and $\mathbf{D^{'}BD^{'}}$ is
positive semi-definite.
Since $\mathbf{D^{'}BD^{'}}\neq\mathbf 0$, it must have at least one non-zero eigenvalue
 $\lambda>0$ with a corresponding eigenvector $\mathbf{v}$.
 Then,
\begin{eqnarray}
\frac{\partial^{2} h_1}{\partial t^{2}}&=&-\mathbf{Tr}\left(\mathbf{(D^{'}BD^{'})(BWCB)}\right)\nonumber\\
&\leq& -\lambda \mathbf{Tr}\left(\mathbf{v}^{T}\mathbf{v}(\mathbf{BWCB})\right)\nonumber\\
&= & -\lambda \mathbf{Tr}\left(\mathbf{v}^{T}(\mathbf{BWCB})\mathbf{v}\right)\nonumber\\
&=&-\lambda \mathbf{Tr}\left(\mathbf{v}^{T}(\mathbf{B(WC)}^{\frac{1}{2}}\mathbf{(WC)}^{\frac{1}{2}}\mathbf{B})\mathbf{v}\right)\nonumber\\
&=&-\lambda\|\mathbf{(WC)^{\frac{1}{2}}Bv}\|^2<0\nonumber.
\end{eqnarray}

Next we prove that the objective function in \eqref{EQ:equivalent_problem} is strictly convex in $\mathbf{W}_k$. The first summation in \eqref{EQ:equivalent_problem} is linear in $\mathbf{W}_k$'s and does not change the strict convexity. Moreover, the objective function in \eqref{EQ:equivalent_problem} is decomposable over $\mathbf{W}_k$. Hence, to accomplish the proof of lemma \ref{lemma: str_concavity} we just need to prove the strict convexity of $-\log \det (\mathbf{W}_k)$ in $\mathbf{W}_k$. For notational simplicity, we drop the index $k$, and prove the strict convexity of $-\log \det (\mathbf{W})$ along any feasible direction within the set of positive-definite matrices. Let  $\mathbf{G}$ be a feasible direction and $t$ be a positive scalar such that $\mathbf{W}+t\mathbf{G}>0$. Then we define 
a one-dimensional parametrization of $-\log \det (\mathbf{W})$ along the direction $\mathbf{G}$
\begin{align}
h_2(t)=-\log \det (\mathbf{W}+t\mathbf{G}).\nonumber
\end{align}
Using properties of the determinant function and the fact that $\mathbf{W}$ is positive-definite, we have
\begin{eqnarray*}
h_2(t)&=&-\log \det (\mathbf{W}^{1/2}(\mathbf{I}+t\mathbf{W}^{-1/2}\mathbf{G}\mathbf{W}^{-1/2})\mathbf{W}^{1/2}) \nonumber\\
&=&-\log \det (\mathbf{W})-\log \det (\mathbf{I}+t\mathbf{W}^{-1/2}\mathbf{G}\mathbf{W}^{-1/2})\nonumber\\
&=&-\log \det(\mathbf{W})-\sum_{i}\log (1+t\lambda_i),\nonumber
\end{eqnarray*}
where $\lambda_i$'s are the eigenvalues of $\mathbf{W}^{-1/2}\mathbf{G}\mathbf{W}^{-1/2}$ and the last step of the above procedure is due to the fact that eigenvalues of $\mathbf{I}+\mathbf{X}$ are one plus the eigenvalues of $\mathbf{X}$.
Obviously for any value of $\lambda_i$ the function $-\log (1+t\lambda_i)$ is convex with respect to $t$ and for any non-zero $\lambda_i$ is strictly convex in $t$. Since $\mathbf{G}$ is non-zero and $\mathbf{W}$ is positive-definite, it follows that there exists at least one non-zero $\lambda_i$ which means that $-\sum_i\log(1+t\lambda_i)$ is strictly convex. Thus, $h_2(t)$ is strictly convex in $t$.
\end{proof}

\begin{lemma}
\label{lemma:Stationarity}
Let $\{(\mbQ_k^*,\mbW_k^*)\}_{k=1}^K$ be a stationary point of \eqref{EQ:equivalent_problem}, then the point $\{\mbQ_k^*\}_{k=1}^K$ is a stationary point of \eqref{EQ:WSRM_Original}. Conversely, if $\{\mbQ_k^*\}_{k=1}^K$ is a stationary point of \eqref{EQ:WSRM_Original}, then $\{(\mbQ_k^*,\mbW_k^*)\}_{k=1}^K$ is a stationary point of \eqref{EQ:Weighted_Sum_Rate}, where $\mbW_k^* \triangleq g_k(\mbQ^*)^{-1}$, $k=1,2,\ldots, K$.
\end{lemma}
\begin{proof}
Let us use ${\psi}_1(\mbQ,\mbW)$  and $\psi_2(\mbQ)$ to denote the objective functions of \eqref{EQ:equivalent_problem} and \eqref{EQ:Weighted_Sum_Rate} respectively, i.e.,
\begin{align}
\psi_1(\mbQ,\mbW) &\triangleq \sum_{k=1}^K \alpha_k \left(\mbW_k g_k(\mbQ)\right) - \sum_{k=1}^K \log \det (\mbW_k),\nonumber\\
\psi_2(\mbQ) &\triangleq \sum_{k=1}^K \log \det (g_k(\mbQ)). \nonumber
\end{align}
Suppose $\{\mathbf{Q}_k^*,\mathbf{W}_k^*\}_{k=1}^K$ is
a stationary point of \eqref{EQ:equivalent_problem}. Since the constraints in \eqref{EQ:equivalent_problem} are separable in the variables,  we have
\begin{align}
&\tr \left(\nabla_{\mbQ_k} \psi_1(\mbQ^*,\mbW^*)^T (\mbQ_k - \mbQ_k^*)\right) \geq 0, \quad  k=1,2,\ldots,K,\label{EQ:Opt_Cond_Q}\\
&\tr \left(\nabla_{\mbW_k} \psi_1(\mbQ^*,\mbW^*)^T (\mbW_k - \mbW_k^*)\right) \geq 0, \quad k=1,2,\ldots,K,\label{EQ:Opt_Cond_W}
\end{align}
for any feasible point $\{\mbQ_k,\mbW_k\}_{k=1}^K$.
By taking the gradient of $\psi_1(\cdot,\cdot)$ with respect to $\mbW_k$ and further simplifying \eqref{EQ:Opt_Cond_W}, we get
\begin{align}
\tr \left[ (g_k(\mbQ^*) - (\mbW_k^*)^{-1} ) (\mbW_k - \mbW_k^*)\right] \geq 0. \nonumber
\end{align}
Since this inequality
holds for any $\mbW_k$, it follows that
\begin{align}
\mbW_k^* = g_k(\mbQ^*)^{-1}, \quad \forall\ k=1,2,\ldots,K. \label{EQ:Opt_W_lemma}
\end{align}
Fix any index $\ell$ and let us use $q_{{m,n}}$ to denote the $(m,n)$-th entry in $\mbQ_{\ell}$. Differentiating using 
chain rule, we obtain
\begin{align}
\frac{\partial \psi_1}{\partial q_{{m,n}}}\,\vline\,_{_{_{(\mbQ^*,\mbW^*)}}} &= \sum_{k=1}^K \alpha_k \tr \left(\mbW_k^* \,
\frac{\partial g_k(\mbQ)}{\partial q_{{m,n}}}\right)\,\vline\,_{_{_{(\mbQ^*,\mbW^*)}}}\nonumber\\
&=  \sum_{k=1}^K \alpha_k \tr \left(g_k(\mbQ^*)^{-1} \,
\frac{\partial g_k(\mbQ)}{\partial q_{{m,n}}}\right)\,\vline\,_{_{_{(\mbQ^*,\mbW^*)}}}\nonumber\\
&= \frac{\partial \psi_2}{\partial q_{{m,n}}}\,\vline\,_{_{_{\mbQ^*}}}, \nonumber
\end{align}
where the second equality follows from \eqref{EQ:Opt_W_lemma}. This further implies that
\begin{align}
 \tr \left(\nabla_{\mbQ_k} \psi_2(\mbQ^*)^T (\mbQ_k - \mbQ_k^*)\right) = \tr \left(\nabla_{\mbQ_k} \psi_1(\mbQ^*,\mbW^*)^T (\mbQ_k - \mbQ_k^*)\right) \geq 0 \quad \forall\ k=1,2,\ldots,K,
\end{align}
which guarantees the stationarity of the point $\{\mbQ_k^*\}_{k=1}^K$ for \eqref{EQ:Weighted_Sum_Rate}. Furthermore, since the objective function in \eqref{EQ:Weighted_Sum_Rate} and the objective function in \eqref{EQ:WSRM_Original} only differ in sign, the stationarity of $\{\mbQ_k^*\}_{k=1}^K$ in \eqref{EQ:Weighted_Sum_Rate} is equivalent to the stationarity of $\{\mbQ_k^*\}_{k=1}^K$ for \eqref{EQ:WSRM_Original}. To prove the converse, we can define $\mbW_k^* = g_k(\mbQ^*)^{-1}$, $k=1,2,\ldots,K$, and simply reverse the above argument to show that \eqref{EQ:Opt_Cond_Q} and \eqref{EQ:Opt_Cond_W} hold. This further implies the stationarity of the point $\{(\mbQ_k^*,\mbW_k^*)\}_{k=1}^K$ for \eqref{EQ:equivalent_problem}.
\end{proof}

\end{document}